\documentclass{article}
\usepackage{ijcai25}

\usepackage{times}
\usepackage{soul}
\usepackage{url}
\usepackage[hidelinks]{hyperref}
\usepackage[utf8]{inputenc}
\usepackage[small]{caption}
\usepackage{graphicx}
\usepackage{amsmath}
\usepackage{amsthm}
\usepackage{booktabs}
\usepackage{algorithm}
\usepackage{algorithmic}
\usepackage[switch]{lineno}
\usepackage{multirow}
\usepackage{enumitem}
\usepackage{newfloat}
\usepackage{listings}
\usepackage{mathrsfs}
\usepackage{amssymb}
\usepackage{arydshln}

\title{SepALM: Audio Language Models Are Error Correctors for Robust Speech Separation}

\author{
Zhaoxi Mu
\and
Xinyu Yang\thanks{Corresponding author}
\And
Gang Wang\\
\affiliations
Xi'an Jiaotong University\\
\emails
\{wsmzxxh, wanggang911\}@stu.xjtu.edu.cn,
yxyphd@mail.xjtu.edu.cn
}

\begin{document}

\maketitle

\begin{abstract}
    While contemporary speech separation technologies adeptly process lengthy mixed audio waveforms, they are frequently challenged by the intricacies of real-world environments, including noisy and reverberant settings, which can result in artifacts or distortions in the separated speech. To overcome these limitations, we introduce SepALM, a pioneering approach that employs audio language models (ALMs) to rectify and re-synthesize speech within the text domain following preliminary separation. SepALM comprises four core components: a separator, a corrector, a synthesizer, and an aligner. By integrating an ALM-based end-to-end error correction mechanism, we mitigate the risk of error accumulation and circumvent the optimization hurdles typically encountered in conventional methods that amalgamate automatic speech recognition (ASR) with large language models (LLMs). Additionally, we have developed Chain-of-Thought (CoT) prompting and knowledge distillation techniques to facilitate the reasoning and training processes of the ALM. Our experiments substantiate that SepALM not only elevates the precision of speech separation but also markedly bolsters adaptability in novel acoustic environments.
\end{abstract}

\section{Introduction}

Speech separation, also known as the cocktail party problem, involves isolating individual speech sources from a mixture of audio signals. Prevailing state-of-the-art techniques in speech separation are predicated on a time-domain dual-path methodology \cite{LuoCY20,SubakanRCBZ21,MuYZ23}, characterized by an encoder-dual-path separation network-decoder framework that adeptly manages lengthy mixed audio waveforms. Despite these advancements, existing methods falter when tasked with separating speech from recordings captured in complex real-world acoustics, such as those rife with noise and reverberation, where residual artifacts or distortions persist in the output. Recent endeavours have sought to surmount this challenge through multi-stage processing \cite{LiLLYZL21,MuYYZ23,NeriB23} and the use of generative models \cite{ChenML20,wang2024noise}. Yet, the efficacy of these approaches in practical scenarios is often constrained, predominantly owing to the heterogeneity of real-world interference that transcends the scope of typical training datasets.

\begin{figure*}[th]
    \centering
    \includegraphics[width=0.93\textwidth]{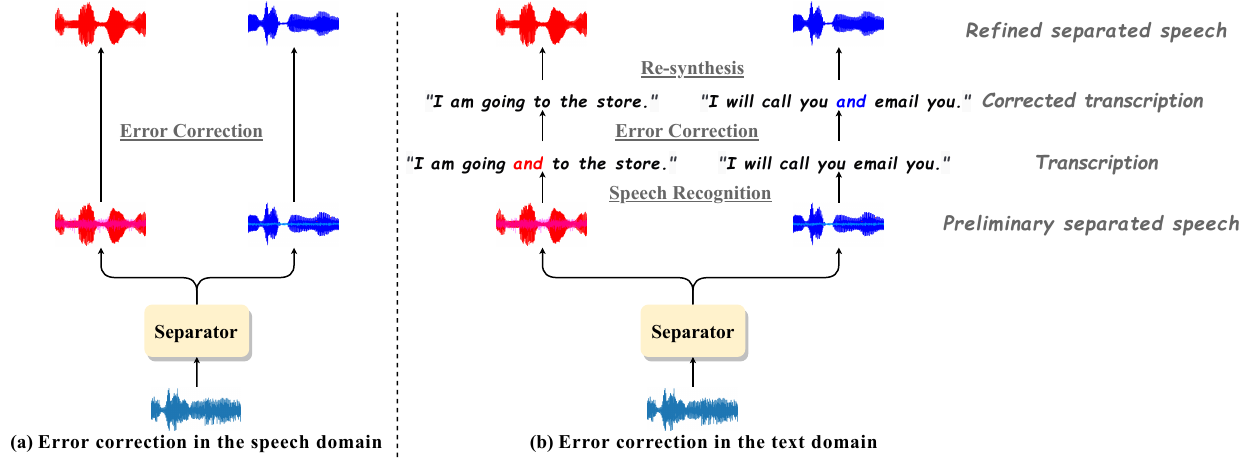}
    \caption{The illustration contrasts (a) the conventional method of error correction for preliminary separated speech in the audio domain, and (b) our proposed method of error correction for preliminary separated speech in the text domain.}
    \label{fig1}
\end{figure*}

To tackle this issue, we propose a novel approach that involves correcting and re-synthesizing preliminary separated speech, which may contain artifacts and distortions, within the text domain, as depicted in Figure \ref{fig1}. The underlying insight behind our decision to perform error correction in the text domain rather than the conventional speech domain is that speech is \textit{high-resolution} data, while text is \textit{low-resolution} data, thus facilitating error correction in a simpler form. Additionally, this textual correction is less susceptible to various types of noise, thereby enhancing the model's adaptability to novel acoustic disturbances.

To obtain transcriptions of the preliminary separated speech and to correct these transcriptions, we propose leveraging the capabilities of large language models (LLMs). These models have recently showcased exceptional proficiency in logical reasoning and language generation, leading to significant achievements and swift progress across various natural language processing tasks \cite{abs-2303-08774,abs-2302-13971}. Trained on extensive textual corpora, LLMs exhibit a robust grasp of world knowledge and nuanced contextual understanding. This has prompted exploration into the application of LLMs for automatic speech recognition (ASR) \cite{abs-2309-10917,abs-2307-11795} and the subsequent refinement of ASR outputs \cite{RadhakrishnanYK23,0075HYSCS23}. However, we have identified several challenges associated with employing a cascade of ASR models and LLMs for error correction. Initially, LLMs can only assess potential errors based on the context provided by the N-best hypotheses decoded by ASR models, lacking the capacity to perceive and utilize the original speech information. This limitation can lead to grammatically coherent yet contextually inaccurate outcomes \cite{abs-2405-10025}. Furthermore, the cascade of ASR models and LLMs escalates computational expenses and the risk of error accumulation, while incompatibilities between the models can complicate optimization and diminish expressive capability.

In light of these insights, we have turned to audio language models (ALMs) \cite{abs-2308-12792}, which are derived from LLMs and exhibit remarkable capacity in processing and generating both textual and auditory information. We advocate for a streamlined, end-to-end error correction strategy utilizing a single ALM. This approach presents two key benefits. First, it capitalizes on the ALM's cross-modal competencies to incorporate original speech data, thereby enhancing the correction process. Second, by integrating a solitary ALM, we alleviate the complexity associated with optimization and inference. To bolster the precision of our single ALM and mitigate potential hallucinations, we have adopted Chain-of-Thought (CoT) prompting \cite{Wei0SBIXCLZ22}. This technique bifurcates the error correction into two phases, enhancing the ALM's reasoning capabilities. Given the paucity of annotated data, we have crafted a knowledge distillation technique that harnesses a pre-trained ASR model as a teacher to direct the ALM's training. Our empirical findings suggest that this CoT strategy rivals the efficacy of more intricate cascaded systems.

Upon securing the corrected transcription, we utilize it alongside the preliminary separated speech to re-synthesize the refined speech, diverging from the conventional use of transcription as a conditional input for speech separation. This decision stems from the recognition that straightforward feature fusion techniques that incorporate transcription information as a condition may precipitate modality imbalance issues \cite{abs-2404-12725}, where the textual modality is prone to ineffectiveness. We experimented with two speech synthesis methods based on the neural codec language model\footnote{This can also be referred to as the audio language model. However, for the sake of differentiation from the previously referenced audio language model, we term this the speech synthesis model based on the neural codec language model.}: autoregressive (AR) generation \cite{BorsosMVKPSRTGTZ23,abs-2301-02111} and non-autoregressive (NAR) masked generation \cite{KharitonovVBMGP23,abs-2305-09636} to resynthesize the refined speech. In our approach, the transcription and the preliminary separated speech are treated equally, being encoded into tokens that serve as inputs to the language model, thereby mitigating the modality imbalance issue. Additionally, we capitalize on the generative model's ability to effectively learn the distribution of clear speech as a refined prior, enhancing the model's generalization capacity when confronted with novel acoustic disturbances.

Due to the lack of phase information during the re-synthesis of refined speech, the synthesized signal may undergo phase shifts over time, which can diminish the precision of metrics evaluated on a time-sample basis, such as the scale-invariant signal-to-noise ratio (SI-SNR) \cite{RouxWEH19}. To mitigate this issue, we implement a realignment process for the refined speech against the preliminary separated speech within the time-frequency domain.

Our contributions are summarized as follows:
\begin{itemize}
\item We propose a novel speech separation paradigm that harnesses the prowess of the audio language model to rectify and re-synthesize preliminary separated speech within the text domain, thereby enhancing the system's ability to separate noisy mixed audio.
\item We advocate for a streamlined, single ALM-based end-to-end error correction mechanism, circumventing the error accumulation and optimization challenges inherent in the traditional sequential integration of ASR models with LLMs. To further enhance the ALM's reasoning and training processes, we have developed techniques of CoT prompting and knowledge distillation.
\item We utilize a speech synthesis method based on the neural codec language model to re-synthesize the refined speech, effectively neutralizing modality imbalance issues and bolstering the model's generalization capability. Furthermore, we incorporate time-frequency domain alignment techniques to resolve phase shift issues, thereby markedly elevating objective evaluation metrics.
\end{itemize}

\section{Related Work}


Recent advancements have led to significant improvements in speech separation performance \cite{LuoCY20,SubakanRCBZ21}. Despite these strides, existing models frequently struggle to maintain separation quality under more demanding acoustic scenarios, such as the presence of more intrusive noise types or diminished signal-to-noise ratios. To surmount this challenge, some studies have put forth generative correction techniques to refine the separated speech further \cite{abs-2301-10752,ErdoganWCBTZH23,abs-2305-05857,wang2024noise}. The underpinning logic of these techniques is that generative models aim to learn the prior distribution of data, can approximate intricate data distributions, and typically exhibit superior generalization capabilities, resulting in the production of more natural and higher-quality samples. Notable approaches similar to ours include Separate and Diffuse \cite{abs-2301-10752} and TokenSplit \cite{ErdoganWCBTZH23}. The method proposed in Separate and Diffuse \cite{abs-2301-10752} amalgamates the strengths of deterministic models with those of stochastic generative models to augment the efficacy of speech separation and perform linear combinations in the time-frequency domain. In TokenSplit \cite{ErdoganWCBTZH23}, the authors propose a scheme for predicting enhanced audio tokens derived from the speech separated by conventional speech separation models and their transcriptions, aiming to eliminate distortions and artifacts present in the separation estimates.

\section{Methodology}

\begin{figure}[t]
    \centering
    \includegraphics[width=0.41\textwidth]{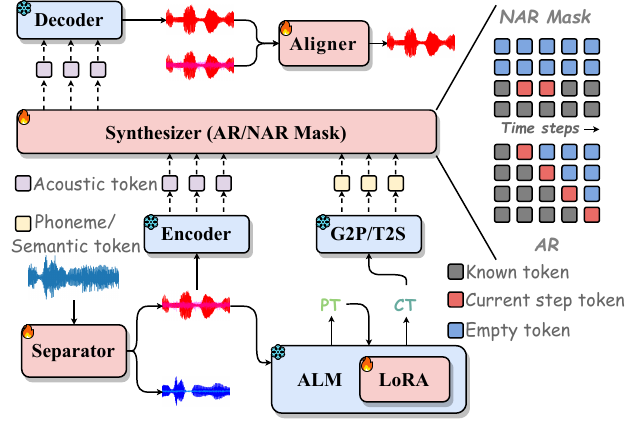}
    \caption{The structural framework of our proposed SepALM. For clarity, the diagram illustrates the processing flow for a single separated speech signal. In the diagram, \textit{`PT'} signifies Preliminary Transcription, and \textit{`CT'} denotes Corrected Transcription.}
    \label{fig2}
\end{figure}

\subsection{Overall Framework}
In this section, we introduce SepALM, a cutting-edge system designed to separate mixed speech and re-synthesize target speech in complex acoustic environments. As depicted in Figure \ref{fig2}, SepALM consists of four principal components: 1) a separator for the preliminary separation of mixed speech; 2) a corrector for error correction of the preliminary separated speech within the low-resolution text domain; 3) a synthesizer that refines the preliminary separated speech by re-synthesizing it based on the corrected text transcription; and 4) an aligner that carries out phase compensation and alignment on the refined speech. Each component will be elaborated upon in the subsequent sections.

\subsection{Separator}

Given a noisy mixed speech signal $x \in \mathbb{R}^{T}$, our goal is to estimate $C$ individual speech sources, denoted as $s^{(i)} \in \mathbb{R}^{T}$ for $i=1,2,\dots,C$. The mixed signal $x$ can be expressed as:
\begin{equation}
x=\sum_{i=1}^C s^{(i)}+n 
\end{equation}
Here, $n \in \mathbb{R}^{T}$ signifies the background noise component, $T$ represents the total number of data points in the signal, and $C$ corresponds to the count of individual speech sources. For the sake of simplicity and without generality being compromised, we assume a scenario with $C=2$ sources.

We initially engage the widely used time-domain dual-path speech separation network, SepFormer \cite{SubakanRCBZ21}, as the separator. This network executes preliminary separation on the noisy mixed speech, yielding preliminary estimates $\hat{s}^{(i)}=f_{sep}(x)\text{ for } i\in {1,2}$. The separator $f_{sep}$ is trained to directly minimize the discrepancy between the estimated signals $\hat{s}$ and the true signals $s$. We utilize the SI-SNR loss to optimize the separator's parameters. However, this conventional speech separation approach frequently encounters issues such as over-suppression or under-separation, particularly in noisy environments. These issues can introduce distortions and artifacts in $\hat{s}$. Consequently, additional corrective measures and refinement processes are warranted to enhance the quality of the preliminary separated speech.

\subsection{Corrector}

In lieu of the prevalent two-stage GER paradigm that incorporates ASR models and LLMs, we posit that employing a singular ALM can yield comparable GER efficacy while simultaneously curtailing the model's computational inference expenses and potential for error accumulation. This hypothesis is substantiated in the experimental section of our study. We harness the pre-trained and fine-tuned ALM, SpeechGPT\footnote{https://huggingface.co/fnlp/SpeechGPT-7B-cm}, to rectify the preliminary estimates $\hat{s}$ within the low-resolution text domain. SpeechGPT undergoes training by segmenting speech into 1,000 HuBERT units and integrating them into the LLaMA-7b\footnote{https://huggingface.co/yahma/llama-7b-hf} \cite{abs-2302-13971} tokenizer, succeeded by fine-tuning LLaMA-7b to learn cross-modal mappings. Equipped with this multi-modal capability, we can adapt the correction process to the source speech.

Our corrector, denoted as $f_{\text{cor}}(\hat{s})=\tilde{t}$, can be formulated as a probabilistic model:
\begin{equation}
\tilde{t}=\arg \max\limits_{t} \mathcal{P}_{\text{cor}}(t\mid \hat{s}) 
\end{equation}
Here, $\tilde{t}$ signifies the corrected transcription. To regulate the response quality and mitigate the risk of potential hallucinations within the ALM, we adopt a CoT approach, bifurcating this task into two sub-steps:
\begin{enumerate}[label=(\Roman*)]
\item The ALM first acts as an ASR model to recognize the speech $\hat{s}$, yielding a preliminary transcription $\hat{t}$. This step is mathematically formulated as:
\begin{equation}
  \hat{t}=\arg \max\limits_{t} \mathcal{P}_{\text{asr}}(t\mid \hat{s})
\end{equation}
\item Subsequently, the ALM functions as a GER model to refine the preliminary transcription $\hat{t}$ by leveraging the original speech $\hat{s}$, resulting in the corrected transcription $\tilde{t}$. This process can be represented as:
\begin{equation}
  \tilde{t}=\arg \max\limits_{t} \mathcal{P}_{\text{ger}}(t\mid \hat{t},\hat{s}) 
\end{equation}
\end{enumerate}
This methodical approach ensures a structured refinement of the transcription, enhancing accuracy and reliability. To implement this procedure, we have crafted an instructive prompt template as follows:

\textbf{[Human]:} \textit{Please transcribe the following speech input, and then use the speech input to correct any errors in the transcribed text. You can do it step by step. This is the input: \{speech unit\} \textless eoh\textgreater .}

\textbf{[SpeechGPT]:} \textit{Preliminary transcription: \{preliminary transcription\}. Corrected transcription: \{corrected transcription\} \textless eoa\textgreater .}

Notably, unlike prior studies that rely on N-best hypotheses generated by ASR systems, we focus on the top-1 hypothesis, which carries the highest probability and is generally deemed to be of the highest quality. This greedy decoding strategy avoids heavy beam search decoding and rescoring procedures, thereby rendering the decoding process more expeditious and efficient. Experimental results presented in Table \ref{tab2} corroborate that the ALM with greedy decoding performs comparably to, or even surpasses, the amalgamations of ASR models and LLMs that are deployed with more intricate decoding techniques.

Given the unavailability of ground-truth transcriptions for these two steps, we have devised a knowledge distillation strategy that employs a pre-trained ASR model as a teacher to direct the training of SpeechGPT. Specifically, we engage Whisper \cite{RadfordKXBMS23} to execute greedy decoding on both the preliminary separated speech $\hat{s}$ and the true clean speech $s$, yielding the target transcriptions $\hat{t}^{\ast}$ and $\tilde{t}^{\ast}$, respectively. Overall, the optimization objectives are encapsulated by the following equations:
\begin{equation}
\mathcal{L}_{\text{asr}}=\sum_{n=1}^{N_1}-\log\mathcal{P}(\hat{t}^{\ast}_n\mid \hat{t}^{\ast}_{n-1},\dots,\hat{t}^{\ast}_1,\hat{s})
\end{equation}

\begin{equation}
\mathcal{L}_{\text{ger}}=\sum_{n=1}^{N_2}-\log\mathcal{P}(\tilde{t}^{\ast}_n\mid \tilde{t}^{\ast}_{n-1},\dots,\tilde{t}^{\ast}_1,\hat{t},\hat{s})
\end{equation}
Here, $\mathcal{L}_{\text{asr}}$ and $\mathcal{L}_{\text{ger}}$ signify the cross-entropy losses associated with the ASR and GER steps, respectively. $\hat{t}^{\ast}_n$ and $\tilde{t}^{\ast}_n$ correspond to the $n$-th tokens of the transcriptions $\hat{t}^{\ast}$ and $\tilde{t}^{\ast}$, respectively. $N_1$ and $N_2$ denote the total count of tokens within the transcriptions. Given the substantial scale of the ALM, we employ the parameter-efficient low-rank adaptation (LoRA) technique \cite{HuSWALWWC22} to fine-tune SpeechGPT, thereby reducing computational and memory overhead.

\subsection{Synthesizer}

Upon acquiring the corrected transcription, our objective is to refine the preliminary separated speech utilizing the corrected transcription. We eschew the conventional approach of re-separating the speech conditioned on the transcription \cite{RahimiAZ22}, instead opting for a strategy that involves re-generating the refined speech. This methodology offers dual benefits: First, by affording equal significance to both text and speech as inputs, we circumvent the potential issue of modality imbalance, which might otherwise render the text modality less effective. Second, by employing a generative model to learn the distribution of clear speech as a prior for refinement, rather than training the model to learn a direct mapping from preliminary separated speech to refined speech, we enhance the model's capacity to generalize to novel acoustic environments.

Inspired by recent advances in neural codec language models for speech synthesis \cite{BorsosMVKPSRTGTZ23}, we conceptualize speech synthesis as a conditional language modelling task utilizing neural codec codes, also known as acoustic tokens. Specifically, we utilize a pre-trained neural codec, DAC \cite{KumarSLKK23}, to tokenize each audio sample into discrete acoustic tokens. We subsequently train a decoder-only language model to generate the acoustic tokens $S^{\ast}$ of the refined speech $s^{\ast}$ using either an AR or NAR masked generation manner, conditioned on the corrected transcription $\tilde{t}$ and the preliminary separated speech $\hat{s}$, represented by the acoustic tokens $\hat{S}$. DAC is a convolutional residual vector quantization audio codec that features a $Q$-level quantizer comprising $B$ entries. The preliminary separated speech $\hat{s}$ is discretized and encoded into $\hat{S} \in \mathbb{R}^{2 \times T^{\prime} \times Q}$, where $T^{\prime}$ denotes the length of the downsampled acoustic tokens.

For the AR generation approach, aligning with methods from prior studies \cite{BorsosMVKPSRTGTZ23,abs-2301-02111}, we adopt a two-stage modelling strategy. In the first stage, an AR language model generates the acoustic tokens for the first quantizer in an AR fashion. In the second stage, a NAR language model generates the acoustic tokens for the remaining $Q-1$ quantizers in parallel. This amalgamation of AR and NAR generation strategies effectively balances the fidelity of the synthesized speech with the inference speed. The training process can be represented as:
\begin{equation}
\begin{aligned}
\mathcal{P}(S \mid \hat{S},p)=\prod_{t=0}^{T^{\prime\prime}}\mathcal{P}_{\text{AR}}( S_{t,1} \mid S_{<t,1},\hat{S}_{:,1},p) \\
\times \prod_{i=2}^Q \mathcal{P}_{\text{NAR}}( S_{:,i} \mid S_{:,<i},\hat{S},p)
\end{aligned}
\end{equation}
where $S \in \mathbb{R}^{2 \times T^{\prime\prime} \times Q}$ denotes the acoustic tokens of the true target speech $s$, with $T^{\prime\prime}$ indicating the length of $S$. The notation $S_{:,<i}$ represents the acoustic token layers in $S$ where the layer indices are less than $i$. The terms $\mathcal{P}_{\text{AR}}$ and $\mathcal{P}_{\text{NAR}}$ represent the AR and NAR generation processes, respectively. We utilize a grapheme-to-phoneme (G2P) tool to convert the corrected transcription $\tilde{t}$ into a phoneme sequence $p$. $p$ is then concatenated with the acoustic tokens $\hat{S}$ of the preliminary separated speech to form the prefix tokens.

For the NAR masked generation approach, akin to methods employed in previous studies \cite{KharitonovVBMGP23,abs-2409-00750}, during training, at each time step $t$, we randomly select a subset of tokens from the $i$-th quantizer $S_{:,i}$ of the true target speech $s$'s acoustic tokens $S$ for masking, resulting in $S_{m,i}$. The synthesizer is then trained to generate the complete target acoustic tokens $S_{:,i}$ in a NAR manner, which can be expressed as:
\begin{equation}
\mathcal{P}_{\text{Mask}}( S_{:,i} \mid S_{m,i},S_{:,<i},\hat{S},P)
\end{equation}
We utilize a pre-trained text-to-semantic (T2S) model based on w2v-BERT \cite{ChungZHCQPW21}, sourced from SPEAR-TTS\footnote{https://github.com/lucidrains/spear-tts-pytorch}, to transform the corrected transcription $\tilde{t}$ into semantic tokens $P$. Subsequently, the embeddings of the semantic tokens $P$ are added to the embeddings of the acoustic tokens $\hat{S}$ and $S_{:,<i}$ to serve as the condition.

The optimization of the synthesizer is achieved by minimizing the negative log-likelihood objective, which equates to the cross-entropy loss between the generated acoustic tokens and the true acoustic tokens. Ultimately, the refined speech $s^{\ast}$ is synthesized utilizing the DAC decoder from the generated acoustic tokens $S^{\ast}$.

\subsection{Aligner}

The potential for phase shifts in the output generated by the generation method necessitates a subsequent alignment process to ensure the refined speech $s^{\ast}$ is accurately aligned with the target. To address this, we integrate the preliminary separated speech $\hat{s}$ and apply phase compensation and alignment to the refined speech $s^{\ast}$. Adhering to the approach outlined in \cite{abs-2301-10752}, we align the two estimated speech signals through a linear combination in the time-frequency domain, yielding the final aligned refined speech $\tilde{s}$, which can be expressed as:
\begin{equation}
\tilde{s}=\text{iSTFT}(\alpha_1 \odot \text{STFT}(\hat{s})+\alpha_2 \odot \text{STFT}(s^{\ast}))
\end{equation}
Here, STFT and iSTFT denote the short-time Fourier transform and its inverse, respectively, while $\odot$ represents the Hadamard product. The weighting coefficients $\alpha_1$ and $\alpha_2$ are determined by the aligner $A$:
\begin{equation}
[\alpha_1, \alpha_2]=A(\text{STFT}(\hat{s}),\text{STFT}(s^{\ast}))
\end{equation}
A detailed exposition of the alignment procedure is provided in the technical appendix. In congruence with the preliminary separated speech $\hat{s}$, the aligned refined speech $\tilde{s}$ is also evaluated using the SI-SNR as the objective function. This guides the training of the entire model by minimizing the divergence between $\tilde{s}$ and the ground-truth speech $s$.

\begin{table*}[th]
\centering
\small
\renewcommand{\arraystretch}{1.1}
\setlength\tabcolsep{1pt}  
\begin{tabular}{lrrrrrrrrrrrr}
\toprule
\multicolumn{1}{c}{\multirow{2}{*}{Method}}                              & \multicolumn{4}{c}{Libri2Mix}                                                                                                                            & \multicolumn{4}{c}{WHAM!}                                                                                                                                & \multicolumn{4}{c}{WHAMR!}                                                                                                           \\ \cline{2-13} 
\multicolumn{1}{c}{}                                                     & \multicolumn{1}{c}{SI-SNRi} & \multicolumn{1}{c}{SDRi} & \multicolumn{1}{c}{NMOS} & \multicolumn{1}{c}{SMOS} & \multicolumn{1}{c}{SI-SNRi} & \multicolumn{1}{c}{SDRi} & \multicolumn{1}{c}{NMOS} & \multicolumn{1}{c}{SMOS} & \multicolumn{1}{c}{SI-SNRi} & \multicolumn{1}{c}{SDRi} & \multicolumn{1}{c}{NMOS} & \multicolumn{1}{c}{SMOS}  \\ \midrule
Conv-TasNet \shortcite{LuoM19}                          & $12.1$                                    & $12.5$                                 & $3.23_{\pm 0.13}$                                & $3.01_{\pm 0.21}$                                & $12.7$                                   & $13.2$                                & $3.32_{\pm 0.30}$                                & $3.25_{\pm 0.34}$                                & $8.3$                                    & $7.8$                                 & $2.82_{\pm 0.24}$                                & $3.17_{\pm 0.32}$            \\
DPRNN \shortcite{LuoCY20}                               & $11.3$                                   & $11.6$                                & $3.21_{\pm 0.28}$                                & $3.04_{\pm 0.19}$                                & $13.7$                                   & $14.1$                                & $3.48_{\pm 0.35}$                                & $3.31_{\pm 0.40}$                                & $10.3$                                   & $9.7$                                 & $3.06_{\pm 0.37}$                                & $3.29_{\pm 0.16}$           \\
Wavesplit \shortcite{ZeghidourG21}                      & $15.1$                                   & $15.8$                                & $3.67_{\pm 0.27}$                                & $3.28_{\pm 0.42}$                                & $16.0$                                   & $16.6$                                & $3.50_{\pm 0.40}$                                & $3.34_{\pm 0.38}$                                & $13.2$                                   & $12.2$                                & $3.17_{\pm 0.12}$                                & $3.38_{\pm 0.44}$            \\
SepFormer \shortcite{SubakanRCBZ21}                     & $12.9$                                   & $13.5$                                & $3.52_{\pm 0.37}$                                & $3.35_{\pm 0.11}$                                & $16.4$                                   & $16.7$                                & $3.63_{\pm 0.19}$                                & $3.37_{\pm 0.23}$                                & $14.0$                                   & $13.0$                                & $3.19_{\pm 0.49}$                                & $\mathbf{3.45}_{\pm 0.37}$   \\
MossFormer2 \shortcite{abs-2312-11825}                  & $16.0$                                   & $16.6$                                & $3.78_{\pm 0.34}$                                & $\mathbf{3.48}_{\pm 0.29}$                       & $18.1$                                   & $18.5$                                & $3.88_{\pm 0.13}$                                & $3.52_{\pm 0.31}$                       & $17.0$                                   & $15.9$                                & $3.20_{\pm 0.31}$                                & $3.43_{\pm 0.47}$  \\
MossFormer2\textsuperscript{*} \shortcite{abs-2312-11825}                  & $16.0$                                   & $16.5$                                & $3.81_{\pm 0.27}$                                & $3.45_{\pm 0.42}$                       & $18.2$                                   & $18.6$                                & $3.92_{\pm 0.21}$                                & $\mathbf{3.55}_{\pm 0.52}$                       & $17.2$                                   & $16.0$                                & $3.22_{\pm 0.13}$                                & $3.37_{\pm 0.25}$            \\ \hdashline
$\text{DiffSep}^{\dagger}$ \shortcite{ScheiblerJCBCC23} & $8.9$                                    & $9.5$                                 & $3.60_{\pm 0.13}$                                & $3.08_{\pm 0.26}$                                & $12.4$                                   & $12.9$                                & $3.67_{\pm 0.34}$                                & $3.15_{\pm 0.42}$                                & $9.5$                                    & $8.6$                                 & $3.16_{\pm 0.49}$                                & $3.13_{\pm 0.19}$            \\
$\text{SepALM}^{\ddagger}$\textsuperscript{*} (\textit{AR})                & $17.4$                                   & $17.9$                                & $\mathbf{3.91}_{\pm 0.33}$                       & $3.29_{\pm 0.28}$                                & $\mathbf{18.8}$                          & $\mathbf{19.3}$                       & $\mathbf{4.01}_{\pm 0.14}$                       & $3.43_{\pm 0.33}$                                & $18.1$                                   & $17.2$                                & $\mathbf{3.44}_{\pm 0.54}$                       & $3.36_{\pm 0.29}$            \\
$\text{SepALM}^{\ddagger}$\textsuperscript{*} (\textit{Mask})              & $\mathbf{17.6}$                          & $\mathbf{18.2}$                       & $3.86_{\pm 0.08}$                                & $3.36_{\pm 0.22}$                                & $18.7$                                   & $19.2$                                & $3.98_{\pm 0.13}$                                & $3.40_{\pm 0.26}$                                & $\mathbf{18.2}$                          & $\mathbf{17.4}$                       & $3.37_{\pm 0.45}$                                & $3.38_{\pm 0.17}$            \\ \bottomrule
\end{tabular}
\caption{Performance comparison of SepALM with other state-of-the-art speech separation models on the Libri2Mix, WHAM!, and WHAMR! benchmark datasets. The symbol $\dagger$ denotes methods based on generative models, while $\ddagger$ signifies methods that integrate both discriminative and generative models. \textit{AR} represents AR generation, and \textit{Mask} signifies NAR masked generation. The superscript $*$ signifies training conducted with the combined dataset of three sources.}
\label{tab1}
\end{table*}

\begin{table*}[th]
\centering
\small
\renewcommand{\arraystretch}{1.0}
\setlength\tabcolsep{2pt}  
\begin{tabular}{lrrrrrrrr}
\toprule
\multirow{2}{*}{Method} & \multicolumn{4}{c}{MUSAN}                     & \multicolumn{4}{c}{DEMAND}                    \\ \cline{2-9} 
                        & SI-SNRi $\uparrow$      & PESQi $\uparrow$        & ESTOIi $\uparrow$  & SIM $\uparrow$     & SI-SNRi $\uparrow$      & PESQi $\uparrow$        & ESTOIi $\uparrow$  & SIM $\uparrow$    \\ \midrule
SepFormer \shortcite{SubakanRCBZ21}               & $9.1$           & $0.50$          & $0.20$   & $\mathbf{0.65}$       & $9.9$           & $0.67$          & $0.21$    & $0.66$      \\
$\text{DiffSep}^{\dagger}$ \shortcite{ScheiblerJCBCC23}                 & $8.4$           & $0.41$          & $0.19$   & $0.48$       & $9.5$           & $0.62$          & $0.20$    & $0.55$       \\
$\text{Refiner}^{\ddagger}$ \shortcite{abs-2305-05857}                & $8.8$           & $0.51$          & $0.20$   & $0.53$       & $9.6$           & $0.60$          & $0.20$    & $0.56$      \\
$\text{Fast-GeCo}^{\ddagger}$ \shortcite{wang2024noise}               & $12.3$          & $0.75$          & $0.30$   & $0.58$       & $13.3$          & $0.92$          & $0.29$   & $0.61$        \\
$\text{SepALM}^{\ddagger}$ (\textit{AR})                  & $13.7$ & $0.82$ & $0.38$ & $0.63$  & $\mathbf{14.5}$ & $1.15$ & $0.46$ & $\mathbf{0.67}$ \\
$\text{SepALM}^{\ddagger}$ (\textit{Mask})                  & $\mathbf{13.9}$ & $\mathbf{0.86}$ & $\mathbf{0.39}$ & $0.62$  & $14.4$ & $\mathbf{1.19}$ & $\mathbf{0.47}$ & $0.65$ \\ \bottomrule
\end{tabular}
\caption{Performance comparison of SepALM with other state-of-the-art speech separation models on the MUSAN and DEMAND out-of-domain noise datasets. The symbol $\dagger$ signifies methods based on generative models, while $\ddagger$ marks generative correction methods that integrate both discriminative and generative models.}
\label{tab1.5}
\end{table*}

\begin{table*}[th]
\centering
\small
\renewcommand{\arraystretch}{1.0}
\setlength\tabcolsep{4pt}  
\begin{tabular}{llrrrrrr}
\toprule
Exp. & Method                                  & SI-SNRi $\uparrow$ & SDRi $\uparrow$ & NMOS $\uparrow$ & SMOS $\uparrow$ & WER (\%)$\downarrow$  & RTF $\downarrow$ \\ \midrule
(a)  & Re-separation model                     & $14.5$               & $15.1$            & $3.64_{\pm 0.52}$            & $3.31_{\pm 0.14}$                                & $3.76 / 4.79$   &   $1.83$           \\ \hdashline
(b)  & SepALM (\textit{AR})   & $17.4$               & $17.9$            & $3.91_{\pm 0.33}$            & $3.29_{\pm 0.28}$                                & $3.76 / 4.10$    &  $2.58$           \\ \hdashline
(c)  & SepALM (\textit{Mask}) & $\mathbf{17.6}$      & $\mathbf{18.2}$   & $3.86_{\pm 0.08}$            &  $\mathbf{3.36}_{\pm 0.22}$                      & $3.76 / 4.03$    &   $1.91$ \\
(d)  & \quad - Separator only                          & $13.2$               & $13.8$            & $3.62_{\pm 0.27}$            & $3.35_{\pm 0.12}$                                & $- / 5.68$   &  $\mathbf{0.52}$            \\
(e)  & \quad - Cascaded method                         & $17.4$               & $17.9$            & $3.93_{\pm 0.29}$            & $3.32_{\pm 0.30}$                                & $3.85 / 4.32$   &   $2.95$           \\
(f)  & \quad \quad - \textit{w/o} Fine-tuning   & $16.3$               & $16.8$            & $3.64_{\pm 0.45}$   & $3.28_{\pm 0.45}$                                & $4.56 / 4.96$       &   $2.95$    \\
(g)  & \quad - \textit{w/o} Aligner   & $15.7$               & $16.2$            & $\mathbf{3.98}_{\pm 0.19}$   & $3.34_{\pm 0.42}$                                & $3.76 / \mathbf{3.99}$       &   $1.89$ \\
(h)  & \quad - \textit{w/o} Fine-tuning   & $17.2$               & $17.8$            & $3.79_{\pm 0.24}$   & $3.30_{\pm 0.19}$                                & $3.93 / 4.25$       &   $1.91$    
\\ \bottomrule
\end{tabular}
\caption{Ablation study on the Libri2Mix dataset. The first and second WER values correspond to the word error rates of the corrected transcriptions and the model's output speech, respectively.}
\label{tab2}
\end{table*}

\section{Experiments}

\subsection{Datasets}

To bolster the generalization capability of our model, we train it using a unified dataset comprising noisy mixed datasets WHAM! \cite{WichernAFZMCMR19}, WHAMR! \cite{MaciejewskiWMR20}, and Libri2Mix \cite{cosentino2020librimix}. WHAM! is a noisy version of WSJ0-2mix \cite{HersheyCRW16}, incorporating noise samples recorded in environments such as cafes, restaurants, and bars. Building upon WHAM!, WHAMR! introduces reverberation effects to the speech sources, supplementing the pre-existing noise components. Libri2Mix is constructed by simulating noise data from WHAM! and speech segments from Librispeech \cite{PanayotovCPK15}, and it encompasses two training subsets: train-360 and train-100. All datasets are sampled at a rate of 16 kHz. The model was trained on the complete audio segments from these datasets to produce semantically coherent transcriptions. During training, the audio segments were padded to ensure uniform length. Upon completion of training, we conduct individual evaluations on each dataset's respective test set.

\subsection{Setup}

For the separator, we utilized SepFormer, which comprises 26M parameters. For the corrector, during fine-tuning, we configured the rank of LoRA to 8, integrated LoRA weights into the query, key, value, and output layers of each Transformer block, and trained the newly added LoRA parameters, resulting in a total of 8M trainable parameters. The optimization process was facilitated by the AdamW optimizer with a peak learning rate of $2e^{-4}$. Training proceeded for 5 epochs with a batch size of 128. During inference, we set the maximum sequence length to 1024 and implemented both Top-$k$ and Top-$p$ sampling strategies, with $k$ set to 40 and $p$ to 0.9. The temperature parameter was set to 0.1, and the beam search was configured with a size of 1. We utilized Whisper-Tiny as the teacher model. For the synthesizer, we implemented a Transformer-based model consisting of 12 layers, 16 attention heads, attention dimensions of 1024, and feed-forward network dimensions of 4096, totaling 202M parameters. The AdamW optimizer was also applied here, with a peak learning rate of $5e^{-4}$. For the NAR masked generation method, we set the number of inference steps to 25. During the training of the synthesizer, we employed classifier-free guidance \cite{abs-2207-12598}, randomly discarding prompts with a probability of 0.1. For the neural codec DAC, we configured $B$ and $Q$ to 1024 and 12, respectively. For the aligner $A$, we employed a two-layer convolutional neural network with residual connections, which has 0.13M parameters. For the entire model, the training strategy commenced with the fine-tuning of SpeechGPT, after which its parameters were frozen while the remaining model components were trained. We employed a permutation-invariant loss function throughout the training process.

\subsection{Evaluation Metrics}

For objective evaluation, we utilized reference-based perceptual evaluation metrics, encompassing scale-invariant signal-to-noise ratio improvement (SI-SNRi), signal-to-distortion ratio improvement (SDRi), perceptual evaluation of speech quality improvement (PESQi), and extended short-time objective intelligibility improvement (ESTOIi). To quantify speech intelligibility, we measured the word error rate (WER) of the generated audio when transcribed by ASR. We utilized Whisper-Tiny to perform ASR on the separated speech to assess the transcription accuracy. To establish a benchmark, we also applied Whisper-Tiny to the original clean speech, treating the resulting transcription as the true reference. To measure the retention of the target speaker's voice by our re-synthesis-based approach, we utilized a speaker verification system based on WavLM \cite{ChenWCWLCLKYXWZ22} to calculate the cosine similarity (SIM) between the speaker embeddings of the generated samples and the real audio. Additionally, We report the real-time factor (RTF) for each method when processing 5 seconds of speech on an A100 GPU to compare inference efficiency. For subjective evaluation, we utilized two metrics: the naturalness mean opinion score (NMOS) and the similarity mean opinion score (SMOS), to assess the naturalness and speaker similarity of the separated speech, respectively. A panel of fifteen human evaluators rated 30 randomly selected speech segments on a scale from 1 to 5, where 1 indicated the lowest quality and 5 the highest.

\subsection{Comparison with State-of-the-Art}

We conducted a comprehensive comparison of our proposed SepALM method against various state-of-the-art techniques, as detailed in Table \ref{tab1}. For previously published results, we reference the original data; otherwise, we present our reproduced outcomes. All baseline methods were trained and evaluated on individual datasets by default. The comparative analysis reveals that SepALM, whether based on AR or NAR masked generation, consistently outperforms baseline methods in most evaluation metrics. This highlights the efficacy of our integrated approach of separation, correction, synthesis, and alignment in processing noisy mixed audio.

Furthermore, it is observed that speech separation methods based on generative models, such as DiffSep \cite{ScheiblerJCBCC23}, tend to underperform on objective metrics, including simulated exact reconstruction and signal-to-noise ratio (SNR), as well as on the subjective metric SMOS, which measures speaker similarity. However, these methods excel in the subjective metric NMOS, which assesses naturalness. This discrepancy arises because, under conditions of high distortion, multiple distinct clean samples may yield identical observed samples post-distortion, compounded by the inherent sampling randomness of generative models. Our method combines the strengths of discriminative and generative models, refining the generative output through alignment to mitigate the non-positive definite issue, thus achieving robust performance across both objective and subjective metrics. In addition, the results presented in rows 5 and 6 of Table \ref{tab1} demonstrate that a simple increase in the quantity of training data did not lead to a noticeable improvement in the performance of MossFormer. This finding suggests that the performance gains observed in our method are predominantly attributed to the error correction and re-synthesis processes we employ.

\paragraph{Evaluation on out-of-domain data.} To assess the generalization capability of our method when faced with unseen noise types, similar to the WHAM! dataset, we utilized the test set of WSJ0-2mix as the speech source and noise audio from the MUSAN \cite{SnyderCP15} and DEMAND \cite{HadadHVG14} datasets as the noise source to simulate out-of-domain noisy mixed audio. Noise is introduced by randomly sampling SNR values from a uniform distribution that spans from -6 to +3 dB. All evaluated methods were trained on the Libri2Mix dataset. The results, as depicted in Table \ref{tab1.5}, indicate that compared to other speech separation methods that apply generative correction directly in the speech domain \cite{abs-2305-05857,wang2024noise}, our text-domain error correction approach exhibits superior generalization performance in out-of-domain noise scenarios.

\subsection{Ablation Study}

In this section, we validate the efficacy of each key design within our method through ablation studies. All models were trained on a unified training set comprising the training sets from WHAM!, WHAMR!, and Libri2Mix and evaluated on the Libri2Mix test set.

\paragraph{Ablation study on the corrector.} Initially, we aimed to substantiate the merit of our proposed text-domain error correction by excluding all elements bar the separator. The outcomes are delineated in Exp. (d) of Table \ref{tab2}, revealing that the correction process significantly improves the quality of the separated speech and reduces the WER. To further validate the feasibility of employing a standalone ALM for GER, we devised a two-stage cascaded model for comparison. Specifically, following Whispering-LLaMA \cite{RadhakrishnanYK23}, we deployed Whisper-Tiny to decode the preliminary separated speech into 3-best hypotheses and then employed LLaMA-7B to perform GER on these hypotheses. The corrected transcriptions were subsequently employed to re-synthesize the separated speech. The training regimen mirrored that of our method. As depicted in Exp. (e) of Table \ref{tab2}, our CoT reasoning-based ALM approach rivals the cascaded paradigm using more intricate decoding techniques and even evinces minor advantages in certain metrics, with a lower RTF. This underscores the superior efficiency of our method compared to the more intricate cascaded paradigm. Furthermore, our method remains effective even without fine-tuning the ALM, as evidenced in Exp. (h) of Table \ref{tab2}. This effectiveness is primarily attributed to the CoT prompts we employ, which enhance the zero-shot reasoning capabilities of the ALM. This feature is absent in traditional two-stage cascaded systems, as illustrated in Exp. (f) of Table \ref{tab2}.

\paragraph{Ablation study on the synthesizer.} To verify the efficacy of the synthesis process within our method, we substituted the synthesizer with a re-separation module analogous to the separator, employing the corrected transcriptions as a condition. This is akin to target speech extraction (TSE) methods \cite{ZmolikovaDOKCY23,MuYSY24} under given conditions. Specifically, we employed a G2P tool to convert the text into phoneme sequences, which were subsequently mapped onto a series of learnable embedding vectors. We utilized the audio embeddings from the intermediate layers of the SepFormer as the value and key, while the phoneme embeddings served as the query for feature fusion based on cross-attention. The outcomes are presented in Exp. (a) of Table \ref{tab2}, revealing that our re-synthesis method surpasses the re-separation method. We attribute this superiority to two principal factors. Initially, for source separation tasks, the performance upper bound of generative models is higher than that of deterministic models \cite{abs-2301-10752}, thereby bolstering the model's generalization capabilities. Second, by harnessing a neural codec language model for speech synthesis, we positioned the corrected transcriptions and the preliminary separated speech on an equal footing, effectively neutralizing the modality imbalance issue. Furthermore, comparing the results of Exp. (b) and (c) in Table \ref{tab2} reveals that the NAR masked generation method outperforms the AR generation method in terms of generation quality and inference speed. This superiority is largely due to its bidirectional attention to the full context and its parallel generation approach.

\paragraph{Ablation study on the aligner.} To assess the contribution of the alignment phase, we omitted the aligner from our process and assessed its impact. The findings, as depicted in Exp. (g) of Table \ref{tab2}, indicate that including the aligner markedly improves objective metrics calculated on time samples, such as SI-SNRi. However, it has a minor detrimental effect on the naturalness (NMOS) of the separated speech. Nonetheless, this trade-off is deemed justifiable given the overall enhancement in the quality of the separated speech.

\section{Conclusion}

This paper introduces SepALM, a novel approach that capitalizes on the prowess of audio language models to correct errors in the preliminary separated speech within the text domain and subsequently re-synthesize it. Our method has been shown to significantly enhance the separation of noisy mixed audio. Ablation studies further substantiated the efficacy of each key component of SepALM. Collectively, our findings highlight the potential of the speech separation-correction-synthesis-alignment paradigm within the text domain, offering new avenues for future research on speech separation under more intricate acoustic conditions.

\newpage
\bibliographystyle{named}
\bibliography{ijcai25}

\end{document}